# Why Some Metal Ions Spontaneously Form Nanoparticles in Water Microdroplets? Disentangling the Contributions of Air–Water Interface and Bulk Redox Chemistry

Muzzamil Ahmad Eatoo[a,b,c,d], Nimer Wehbe[e], Najeh Kharbatia[b], Xianrong Guo[e], Himanshu Mishra[a,b,c,d*]

[a]Environmental Science and Engineering (EnSE) Program, Biological and Environmental Science and Engineering (BESE) Division, Water Desalination and Reuse Center (WDRC)

[b]Water Desalination and Reuse Center (WDRC),

King Abdullah University of Science and Technology (KAUST), Thuwal, 23955-6900, Kingdom of Saudi Arabia

[c]Center for Desert Agriculture (CDA), King Abdullah University of Science and Technology (KAUST), Thuwal 23955-6900, Saudi Arabia

[d]Interfacial Lab (iLab), King Abdullah University of Science and Technology (KAUST), Thuwal 23955-6900, Saudi Arabia

[e]Core Labs, King Abdullah University of Science and Technology (KAUST), Thuwal 23955-6900, Saudi Arabia

*Correspondence: himanshu.mishra@kaust.edu.sa

## Abstract

Water microdroplets containing 100 µM $HAuCl_4$ have been shown to reduce gold ions into gold nanoparticles spontaneously. It has been suggested that this chemical transformation is driven by ultrahigh electric fields at the air–water interface, albeit without mechanistic insight. We investigated the fate of several metallic salts in water, methanol, ethanol, and acetonitrile in bulk and microdroplets. This revealed that when $HAuCl_4$ (or $PtCl_4$) is added to bulk water (or methanol or ethanol), metal NPs appear spontaneously. Over time, the nanoparticles grow in bulk, as evidenced by the solutions' changing colors. If the same bulk solution is sprayed pneumatically and collected, the NP size has no significant enhancement. Interestingly, the reduction of metal ions is accompanied by the oxidation of water (or alcohols); however, these redox reactions are minimal in acetonitrile. We establish that the spontaneous reduction of metal ions is (i) not limited to water or gold ions, (ii) not driven by the air-water interface of microdroplets, and (iii) appears to be a general phenomenon for solvents containing hydroxyl groups. These results advance our understanding of liquids in general and should be relevant in soil chemistry, biogeochemistry, electrochemistry, and green chemistry.

## Introduction

Biochemistry, chemistry, and chemical engineering textbooks often limit the role of water in chemical reactions to dissolving/hydrating molecules and ions[1-3]; however, this view is rapidly evolving[4-9]. In the last two decades, experimental and computational studies have revealed that the skin of water – the air-water interface – and water–hydrophobe interfaces, in general, have anomalous features such as enhanced speciation of ions[10, 11] and electrified interfaces[12-15]. Several studies have reported on orders of magnitude faster (e.g., $10^2$–$10^6$×) reaction rates in water microdroplets compared with the bulk phase, thereby expanding the function of aqueous interfaces in chemical transformations beyond heat and mass transfer[5, 16-24]. This exciting field – Microdroplet Chemistry – has emerged as a frontier in chemical science with potential implications for atmospheric aerosol[25]; sea sprays[26]; disease transmission[27, 28]; life's origins[16, 19, 29]; chemistries in clouds, fog, smog, and dew[30-32]; soil processes[33]; green chemistry[18]; hospital disinfection[34]; food industry[35]; and fizzy beverages[36]; also, one may expect similar trends in microbubbles/foam in fermenter broths[37], wastewater treatment[38, 39], and electrochemistry[40, 41].

While much excitement exists, several claims for purely interfacial chemistry being responsible for the rate enhancement have been challenged. This is because a vast majority of reports exploit electrospray ionization mass spectrometry (ESIMS), which entails rapid solvent evaporation, solute concentration, temperature, and electric field gradients, etc., conditions that are far from thermodynamic equilibrium and, therefore, not representative of common systems[20, 23, 42, 43]. For instance, the emergence of superacid chemistry in the microdroplets of mildly acidic water (pH < 4)[44, 45] and the formation of abiotic sugar phosphates in water microdroplets[16], which have been challenged by Mishra & co-workers[46-48] and Wilson & co-workers[49], respectively (for recent reviews on this subject, see Ref.[15, 24, 50, 51]. Controversial reports on microdroplet chemistry are not limited to ESIMS alone. For instance, recently, Zare & co-workers claimed that the air–water interface of microdroplets spontaneously generates $H_2O_2$[52-55], and the reported amounts varied from ~1 ppm or 30 µM (in 2019 for pneumatically sprayed droplets)[52] to 114 µM (in 2020 for condensed droplets)[54] to 180 µM (in 2021 for microdroplets condensed at 50% relative humidity)[53] to 0.3-1.5 µM (in 2022 for sprayed droplets in othe zone-free gas environment, which increased four folds if $O_2$(g) was used to spray water pneumatically)[55]. As for mechanistic insights, the computer simulations by Head-Gordon & co-workers[56, 57] suggested the emergence of instantaneous ultrahigh electric fields at the air–water interface that may drive $H_2O_2$ formation, while the simulations of Ruiz-Lopez and co-workers[50] suggested that the local electric field was smaller at the air–water interface compared to the bulk and emphasized the role of significant fluctuations of the electrostatic potential. Cooks and co-workers have speculated (basing it on Amotz's commentary[58] on a paper by Roke & co-workers on water–oil electrification[12]) that the air–water interface of microdroplets promotes the formation of water radial cations $H_2O^{+*}$ and anions $H_2O^{-*}$, which upon separation during spraying drive redox reactions[24]. Colussi also proposed a similar mechanism whereby spraying yielded a small fraction of oppositely charged droplets, i.e., comprising excess $H_3O^+$ and $OH^-$. These droplets collided to form OH* and H* radicals[59].

Our experimental investigation uncovered that the first two reports from Zare & co-workers, claiming 30–114 -µM $H_2O_2$ formation, were, in fact, reporting artifacts resulting from ambient ozone gas[60, 61]. In their latest paper, Zare & co-workers have acknowledged their mistake[55, 62, 63]. When they repeated their experiments in an ozone-free environment[55, 62, 63] and found ~0.3- 1.5 µM $H_2O_2$(aq) in sprayed microdroplets and, therefore, doubled down on their original claim; also, they noted that the $H_2O_2$(aq) concentration in water microdroplets increases around four times if the nebulizing gas was changed from $N_2$(g) to $O_2$(g). Our latest report demonstrated that this trace $H_2O_2$(aq) formation (i.e., 0.3–1.5 µM) takes place at the water–solid interface and not the air–water interface[64]. This happens as the dissolved oxygen ($O_2$(aq)) gets reduced and the surface gets oxidized, i.e., if $O_2$(aq) is eliminated from the water, $H_2O_2$(aq) is not observed within a detection limit of 50 nM. Notably, the amount of $H_2O_2$(aq) formed at various water–solid interfaces (e.g., stainless steel, Mg, Ti, Si, and $SiO_2$) varies in accordance with the Galvanic series. Our findings thus disprove the mechanisms for the spontaneous $H_2O_2$(aq) formation proposed by Zare and co-workers[27, 52-55, 65, 66], Head-Gordon and co-workers[56, 57], Cooks and co-workers[24], and Colussi[59]. Recently, Nguyen & co-workers[67] and Koppenol & co-workers[68] have also refuted the $H_2O_2$(aq) generation claims in condensed droplets. Given the inconsistent experimental results and multiplicity of explanations for the microdroplet phenomena, a careful and case-by-case assessment of chemistries at the air–water interface and their implications in natural and applied sciences is warranted.

Based on these findings, we investigate another claim from Zare & co-workers on the spontaneous reduction of $Au^{3+}$(aq) to Au nanoparticles (Au NPs) in water microdroplets containing chloroauric acid ($HAuCl_4$, 100 μM concentration)[69]. This phenomenon is also speculated to exclusively occur at the air–water interface of microdroplets due to ultrahigh interfacial electric fields, i.e., without a reducing agent or an electrical voltage required for the bulk water phase. However, it is not entirely clear what the special role of the water interface or the gold ions is in this process or if $H_2O_2$(aq) formation is implicated in this phenomenon also[70]. It has been noted that gold nanoparticles can be formed in bulk water by applying microwaves[71]; others have shown that light can impact the reduction of $Fe^{3+}$ to $Fe^{2+}$ in water[72]. Herein, we present the results of our investigation along the following lines of inquiry:

1. Is the spontaneous reduction of gold in water limited only to water microdroplets, or does it also occur in bulk water? Note: the microdroplets are produced pneumatically.
2. If gold ions get spontaneously reduced inside water, does water get oxidized?
3. What are the effects of gold ion concentration and water pH on the NP formation?
4. Do other metallic ions, e.g., $Pt^{4+}$ and $Fe^{3+}$, also exhibit this behavior?
5. Does water have unique properties, or can other common liquids such as alcohols also drive the spontaneous reduction of metallic ions? If so, what are the reaction byproducts?
6. What are the fundamental molecular mechanisms underlying these chemical transformations?

We observed that when $Au^{3+}$ (or $Pt^{4+}$ or $Fe^{3+}$) was added to water, methanol, or ethanol, they were spontaneously reduced in the bulk phase as well as in water microdroplets. In fact, in this study, whenever NPs formed for a specific salt–solvent system, they formed both in the bulk and microdroplets. In other words, we found no scenario wherein the spontaneous reduction occurred exclusively inside microdroplets but not in bulk.

**Results**

In this work, we used MilliQ Advantage 10 (18 MΩ-cm, 3 ppb) deionized water, albeit the results were identical with HPLC grade water. Microdroplets were obtained via pneumatic nebulization following the experimental setup of Zare et al. [69]. The spray device comprised two concentric capillaries, and the solution was injected through a 0.1-mm-wide inner stainless-steel capillary at a rate of 25 µl/min. The solution was sheared via high-pressure dry $N_2$ gas flowing through the outer concentric capillary at 100 psi, forming a microdroplet stream (Supplementary Fig. S1).

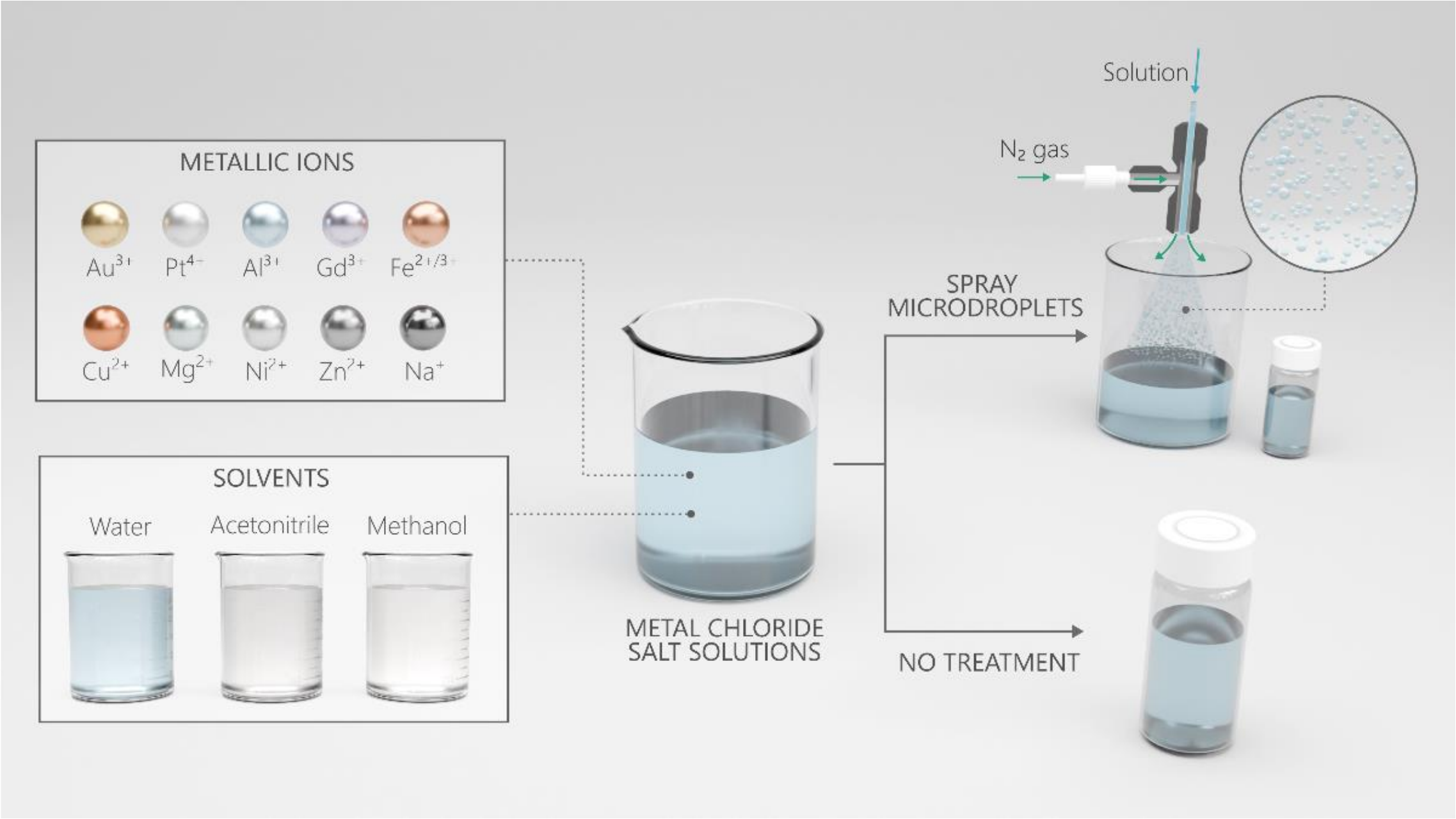

**Fig. 1** Pneumatic spraying setup and various salt solutions.

The effects of the air–water interface on NP formation were analyzed by spraying 100-µM aqueous $HAuCl_4$ solution via pneumatic nebulization, which yielded microdroplets of an average size of ~20 µm (Supplementary Fig. S1). The microdroplets were collected in clean glass vials and transferred into transmission electron microscopy (TEM) grids for TEM imaging and $SiO_2$/Si wafers for scanning electron microscopy (SEM) analysis, which revealed NP formation (Supplementary Fig. S2). Moreover, the transparent mother solution (100-µM aqueous $HAuCl_4$ stored at 22℃ inside a laboratory) changed its color to red in a few (2-3) days and blue in a few (2-3) weeks (Fig. 2(a)). The TEM analysis of samples drawn from the mother solutions revealed the presence of gold NPs (Figs. 2c,d, and S3). Dynamic light scattering (DLS) showed the appearance of a bimodal nanoparticle size distribution in the freshly prepared solutions (Fig. 2(b)). X-ray photoelectron spectroscopy compared the oxidation states of gold in the newly prepared and one-week-old solutions (Fig. 2(e), Methods). The oxidation state of gold in the freshly prepared solutions was +3 and +1, whereas that in the 1-week-old samples was 0. Irrespective of the duration of XPS measurements, an oxidation state of +3 was not recorded; this indicated that the reduction was spontaneous. These results demonstrate that the reduction of gold ions can occur spontaneously in bulk water and is not exclusive to the microdroplet environment.

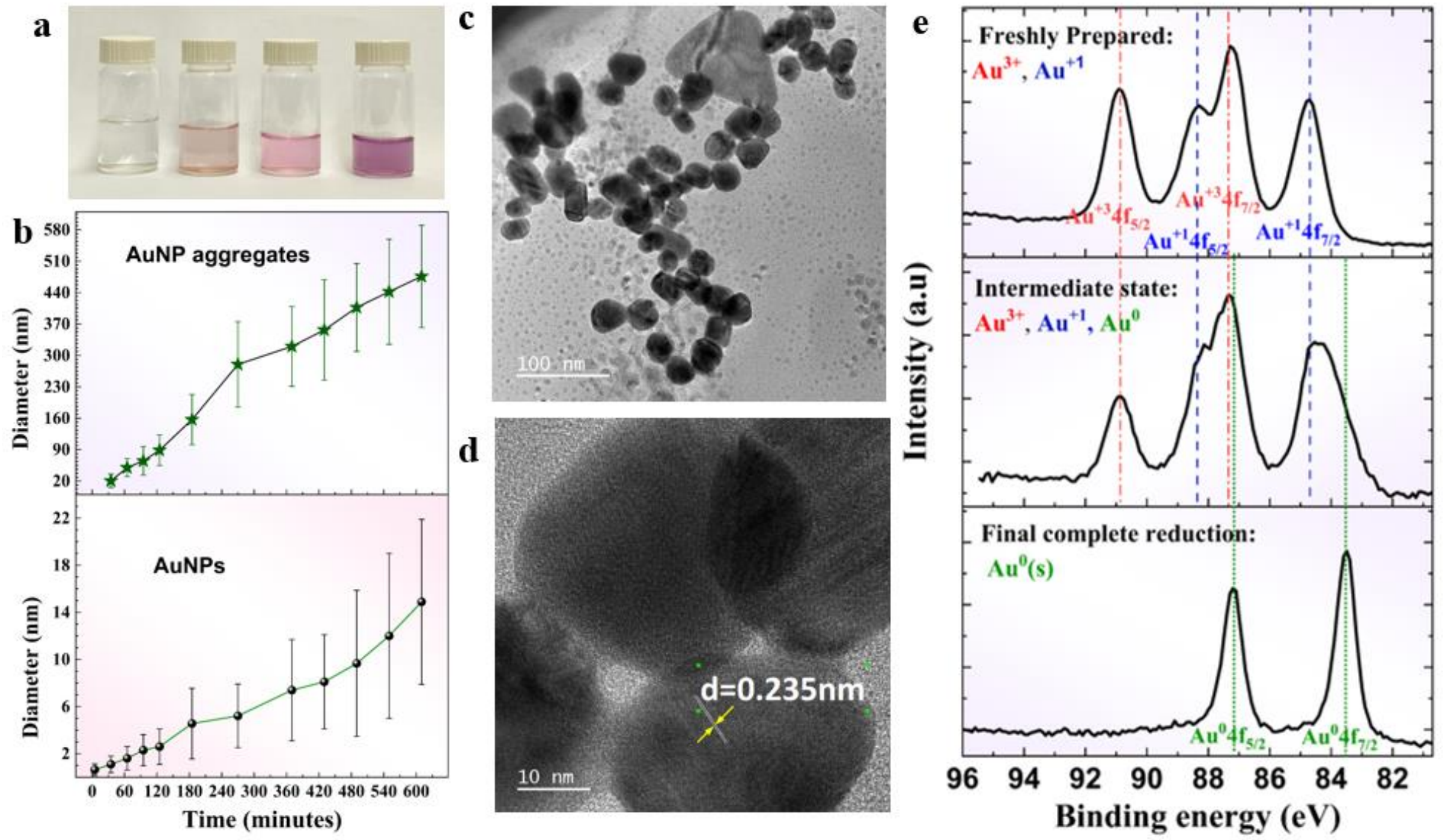

**Fig. 2** Spontaneous formation and growth of AuNPs in bulk 100-µM $HAuCl_4$ in water. **a** Bulk solution changing color due to the nucleation and growth of AuNPs. **b** Formation and growth of particles with time, as measured by DLS. **c** TEM micrograph showing the presence of AuNPs of different sizes. **d** TEM imaging showing an interlayer spacing of ~0.235 nm corresponding to the Au(111) atomic plane of metallic gold Au(0). **e** XPS analysis showing the presence of $Au^{3+}$ and $Au^{+}$ in freshly prepared solution, $Au^{3+}$, $Au^{+}$, and $Au^{0}$ in 1-day-old solution, and $Au^{0}$ in 1-week-old solution. Complete reduction of gold ions to AuNPs after some time.

However, whether the microdroplet environment, i.e., the air–water interface, considerably accelerated the rate of reduction and nanoparticle formation compared with the bulk phase was further investigated. To this end, bulk solutions of the same concentration (100-µM HAuCl4) but different ages, i.e., varying from seconds to minutes to hours, were pneumatically sheared into water microdroplets. The flying microdroplets were then intercepted and collected (in the liquid state, i.e., before complete evaporation) and analyzed via DLS. The observed particle size distributions were compared against those of the mother solution (Fig. S4). Results revealed that the microdroplets afforded no significant enhancement in the particle sizes over the bulk phase, demonstrating that microdroplets do not have any significant effect on the formation or growth of nanoparticles. In summary, the reduction of gold ions in bulk water is spontaneous, and microdroplets are not essential, or the size of microdroplets is not the driving force for the spontaneity of this reduction reaction.

Since the formation of AuNPs from $Au^{3+}$ is a reduction reaction, a parallel oxidation reaction must occur simultaneously. Therefore, to reveal the redox reaction's oxidation half-reaction, we performed solution-state NMR spectroscopy to investigate the possible oxidation products in a liquid state. Gas chromatography (GC) was performed to determine the evolution of the formed gases. NMR spectroscopy revealed the spontaneous formation of $H_2O_2$(aq) in the bulk $HAuCl_4$ solution, and GC revealed the gradual evolution of oxygen ($O_2$) gas in the headspace (Figs. 3a and S5). The formation of $H_2O_2$ was also observed and quantified using HPAK (Fig. 3b). The mechanistic insights into this chemistry are elaborated in the Discussion section.

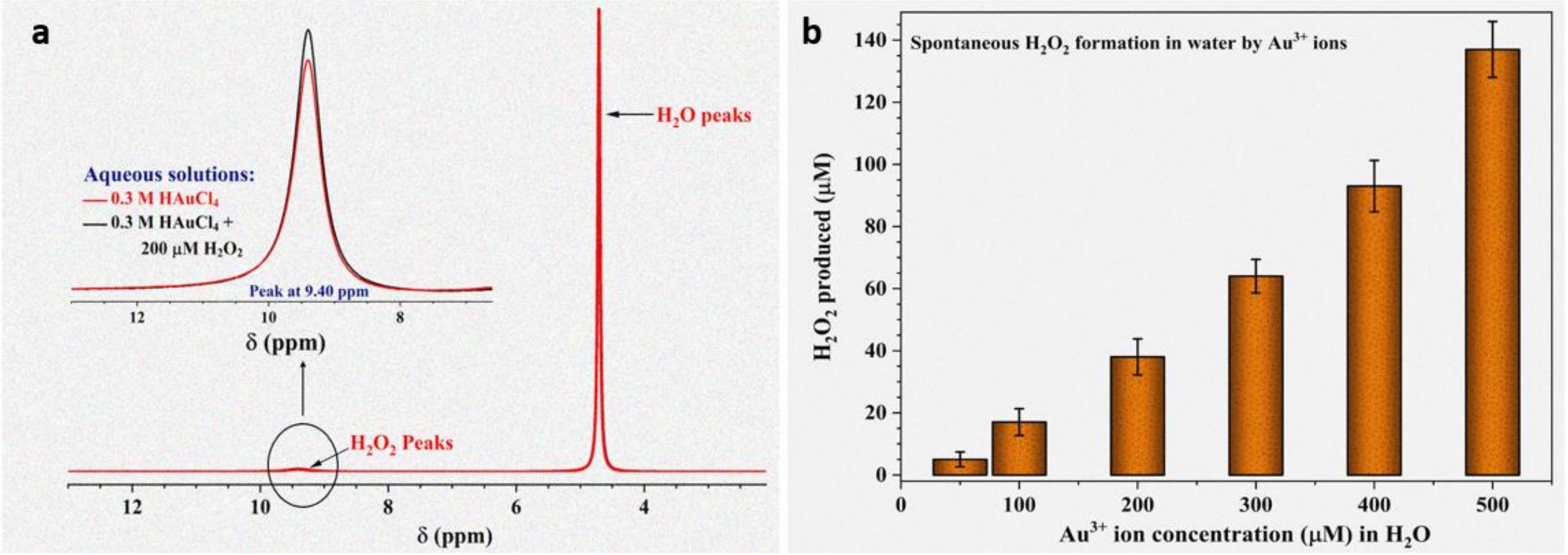

**Fig. 3** Production of $H_2O_2$ during the reduction of $Au^{3+}$ in bulk water. **a** NMR spectroscopy confirming the formation of $H_2O_2$. **b** Quantification of $H_2O_2$ in aqueous $HAuCl_4$ solutions.

The concentration of $H_2O_2$(aq) formed spontaneously during the reduction of gold ions in bulk water increased with $Au^{3+}$ concentration (Fig. 3b), possibly due to the oxidation of $H_2O$ that produced $H_2O_2$ rather than the reduction of $O_2$(aq). $H_2O_2$(aq) formation in $HAuCl_4$ solutions prepared using water with dissolved oxygen and deoxygenated water was examined. We found that the dissolved oxygen had no measurable effect on $H_2O_2$ formation in aqueous $HAuCl_4$ solutions (Fig. S6). Thus, we confirmed the formation of $H_2O_2$(aq) from bulk water as the oxidation half-reaction occurring during the spontaneous formation of Au NPs.

Next, we investigated the effects of solution pH and concentration of $Au^{3+}$ on the formation of AuNPs and corresponding $H_2O_2$ in water. As the solution turned acidic, the formation of the AuNPs and the corresponding production of $H_2O_2$ decreased (Fig. 4a–-d). SEM imaging revealed that when a 200-μM $HAuCl_4$ solution was prepared using deionized water, 0.1-M HCl, or 1-M HCl, the nanoparticles were well-formed (Fig. 4b) or smaller and lesser (Fig. 4c), or nearly absent (Fig. 4d), respectively. This indicates that gold- ions are more stable at lower pH or highly acidic aqueous solutions. The effect of gold ion concentration on the formation of Au NPs and $H_2O_2$ was also examined. Results revealed that gold ions could undergo complete reduction to form Au NPs in relatively diluted solutions (<0.05 M), whereas in concentrated solutions (>0.05 M), $Au^{3+}$ primarily reduced to lower oxidation states (mostly $Au^{+1}$). In both cases, $H_2O_2$ was produced stoichiometrically. XPS also revealed that as the salt concentration increased, the ratio of hydrated ions $Au^{1+}/Au^{3+}$ decreased (Fig. 4e).

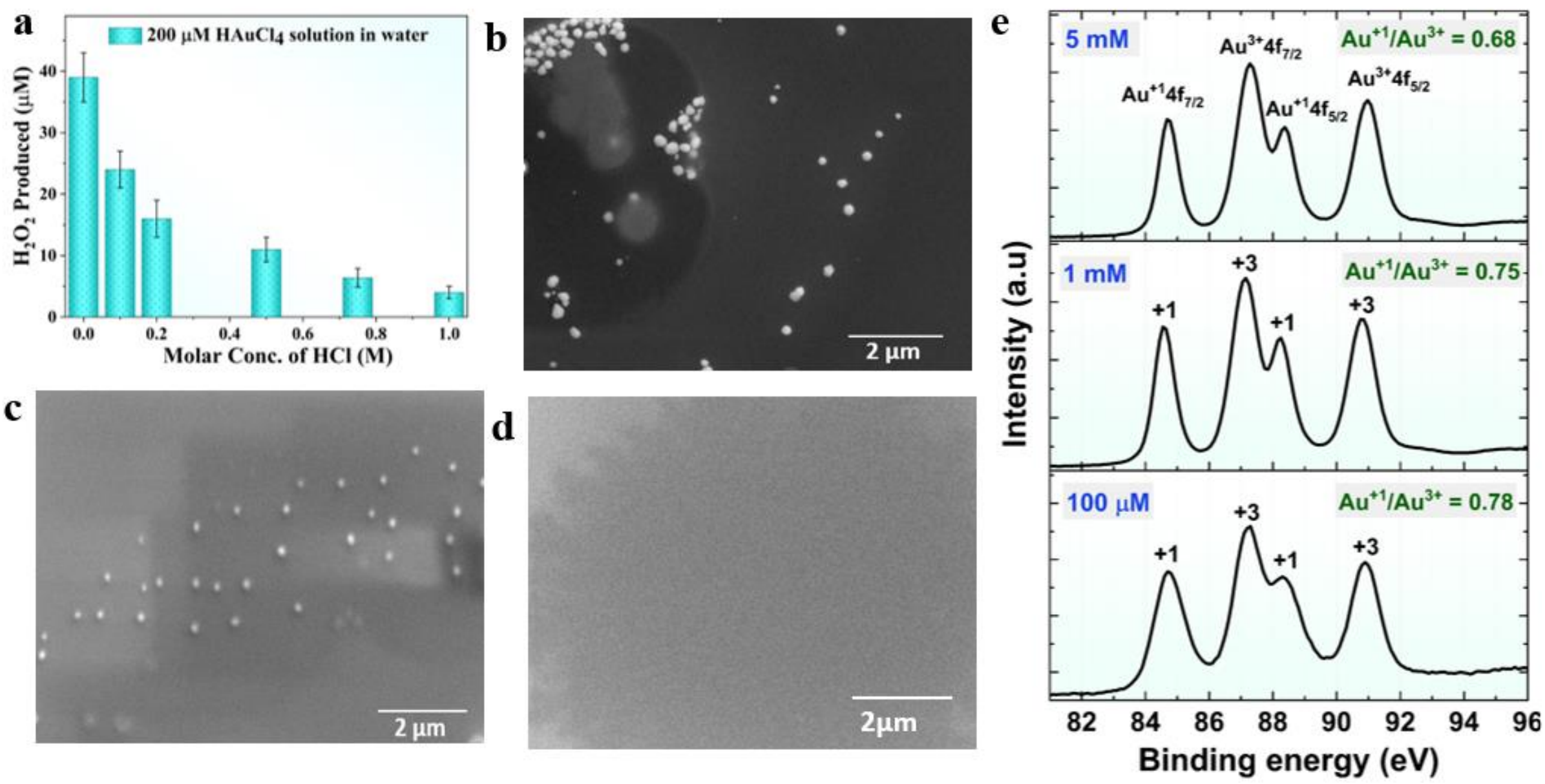

**Fig. 4** Effects of the solution pH and concentration on the reduction of gold ions: **a** Decrease in the rate of $H_2O_2$ production in bulk solutions as a function of HCl concentration (or pH) for a fixed $HAuCl_4$ concentration. **b** SEM micrograph of gold NPs formed in a 200-μM $HAuCl_4$ solution formed with deionized water; **c** aqueous solution formed with 0.1-M HCl; **d** aqueous solution formed with 1M HCl solution. **e** XPS reveals the ratio of $Au^{1+}/Au^{3+}$ in freshly prepared $HAuCl_4$ solutions of varying concentrations (100 μM, 1 mM, and 5 mM).

To further elucidate whether the mechanisms underlying this chemical transformation, i.e., the spontaneous reduction of $Au^{3+}$ in water, is unique or common with other metal ions, a few common metal ions such as $Pt^{4+}$, $Al^{3+}$, $Fe^{3+}$, $Gd^{3+}$, $Fe^{2+}$, $Cu^{2+}$, $Mg^{2+}$, $Ni^{2+}$, $Zn^{2+}$, and $Na^{+}$ were also tested. The reduction of $Au^{3+}$ in other solvents such as methanol, ethanol, and acetonitrile were studied to analyze the uniqueness of water or the air- water interface. We found that two metal cations, $Pt^{4+}$ and $Fe^{3+}$, produced $H_2O_2$ in water (Fig. 5c,d), and the remaining cations did not produce $H_2O_2$. Moreover, considerably higher amounts of $H_2O_2$ were produced by $Au^{3+}$ reduction than those produced via $Pt^{4+}$ and $Fe^{3+}$ reduction.

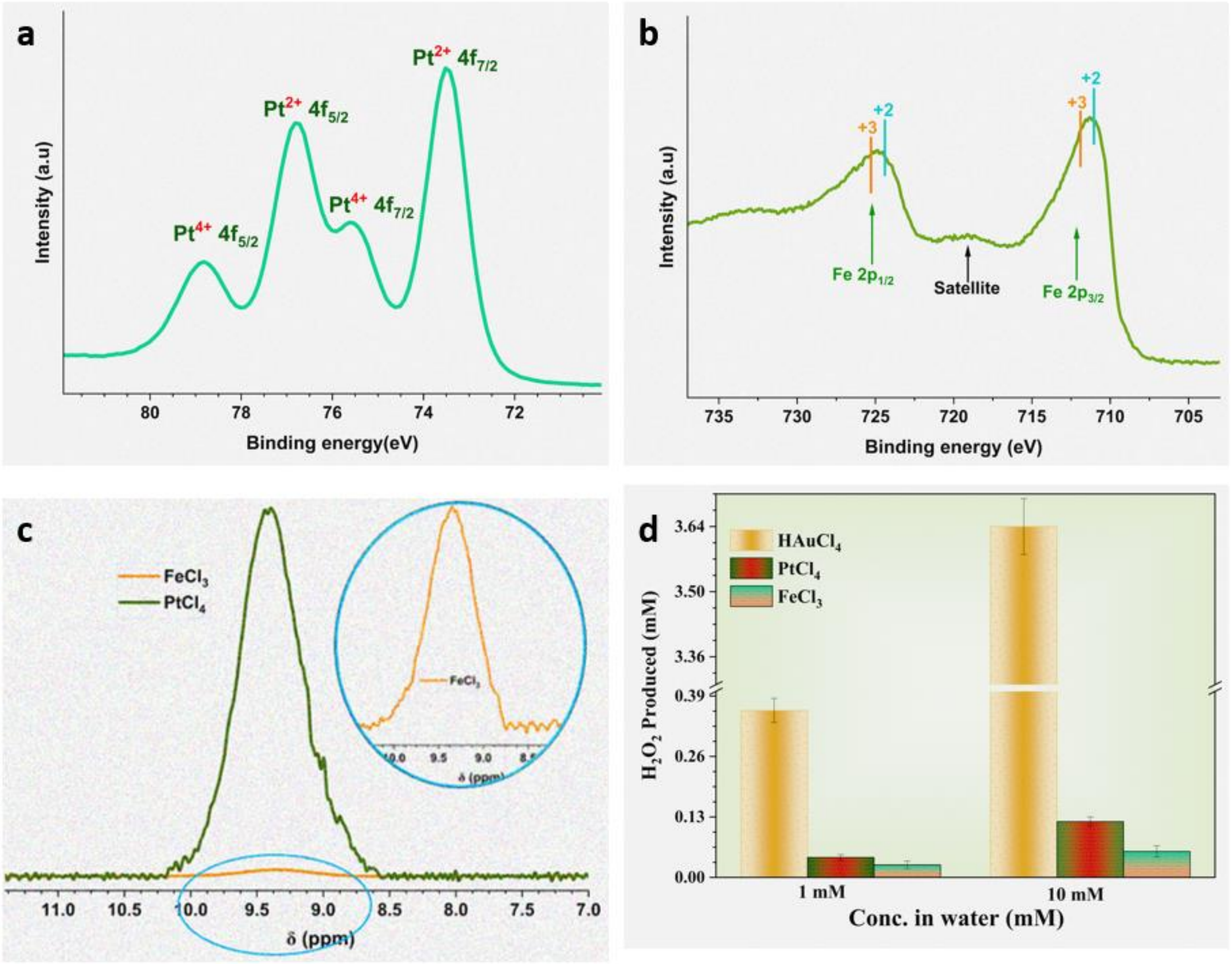

**Fig. 5** Spontaneous reduction of metal ions, besides $Au^{3+}$, in water, which yields $H_2O_2$(aq). XPS results of 100-µM $PtCl_4$ in water showing the reduction of $Pt^{4+}$ to $Pt^{2+}$; Pt NPs observed via TEM images (Fig. S7). **b** XPS results of $FeCl_3$ in water showing the spontaneous reduction of $Fe^{3+}$ ions to $Fe^{2+}$. **c** NMR spectra confirming the formation of $H_2O_2$(aq) in aqueous solutions of $FeCl_3$ and $PtCl_4$ in water (each at a 100-µM initial concentration). **d** Comparison of $H_2O_2$(aq) formed by different concentrations of $Au^{3+}$, $Pt^{4+}$, and $Fe^{3+}$ in water.

XPS analysis confirmed that $Pt^{4+}$ and $Fe^{3+}$ were spontaneously reduced in water similar to $Au^{3+}$, yielding $Pt^{2+}$ and $Fe^{2}$ (Fig. 5a,b), and the formation of Pt NPs was observed by TEM imaging (Fig. S7). Notably, the reduction of $Fe^{3+}$ to its zero-valent form was not observed. In other words, of all the metal ions studied herein, the pure metallic form of NPs was only observed in $HAuCl_4$ and $PtCl_4$ solutions.

## Discussion

This section consolidates all the findings and presents mechanistic insights. We unambiguously established that the spontaneous reduction of $Au^{3+}$ in water forms Au NPs and that this phenomenon is not limited to water microdroplets. This reduction reaction starts in the bulk solution right after its preparation. In fact, spraying bulk solutions and collecting the microdroplets reveals no significant enhancement in the size of AuNPs, suggesting that the microdroplet environment's effects, if any, are minor, such as facilitating solvent evaporation. Based on our experimental results, wherein $Au^{3+}$ in bulk $HAuCl_4$ solutions spontaneously reduce to $Au^+$ ions and AuNPs and drive the formation of $H_2O_2$(aq) (Figs. 2, 3, and S5) and decrease in the solution pH, we suggest the following half-reactions[40]:

| *Reduction half-reactions:* | *E° (volts)* | |
|---|---|---|
| $AuCl_4^-$ (aq) + $3e^- \rightleftharpoons$ Au(*s*) + $4Cl^-$ | 1.002 | (1) |
| $AuCl_4^-$ (aq) + $2e^- \rightleftharpoons AuCl_2^-$ (aq) + $2Cl^-$ | 0.926 | (2) |

*Possible feasible oxidation half-reactions that may drive the above reduction reactions:*

| | | |
|---|---|---|
| $O_2(g) + 2H_2O(l) + 4e^- \rightleftharpoons 4OH^-$ | 0.401 | (3) |
| $H_2O_2 + H^+ + e^- \rightleftharpoons HO^* + H_2O$ | 0.710 | (4) |
| $O_2(g) + 2H^+ + 2e^- \rightleftharpoons H_2O_2$ | 0.695 | (5) |

The standard reduction potentials of equations (1) and (2) are higher or nobler than those of equations (3) and (4); therefore, it is possible that hydroxyl ions ($OH^-$ (aq)) and $H_2O$ may be oxidized to form $H_2O_2$(aq), which later decomposes/oxidizes gradually with time and forms $O_2$(g) (Eq. (5) further reducing gold ions.

The formation of Au NPs in common solvents such as alcohols and acetonitrile were also investigated. We observed that $Au^{3+}$ rapidly reduced in methanol and ethanol than in bulk water (Fig. 6a,b), and it was negligible/sluggish in acetonitrile (Fig. 6a,b). In a complementary experiment, we quantified the amount of the reducing agent (standard hydrogen peroxide, 30%) required to completely reduce a fixed concentration of gold solution (0.1-M $HAuCl_4$) in methanol, water, and acetonitrile; the volume percent of $H_2O_2$ required were $0.4 \pm 0.1$, $1.0 \pm 0.2$, and $35 \pm 1$, respectively. This indicates that Au NP formation is not unique to water. In a methanol solution containing dissolved $HAuCl_4$, we observed that the formation of Au NPs was accompanied by the formation of $H_2O_2$, methylal ($CH_3OCH_2OCH_3$), and dimethyl ether ($CH_3OCH_3$) (Fig. S8). We did not investigate mechanisms underlying these chemical transformations.

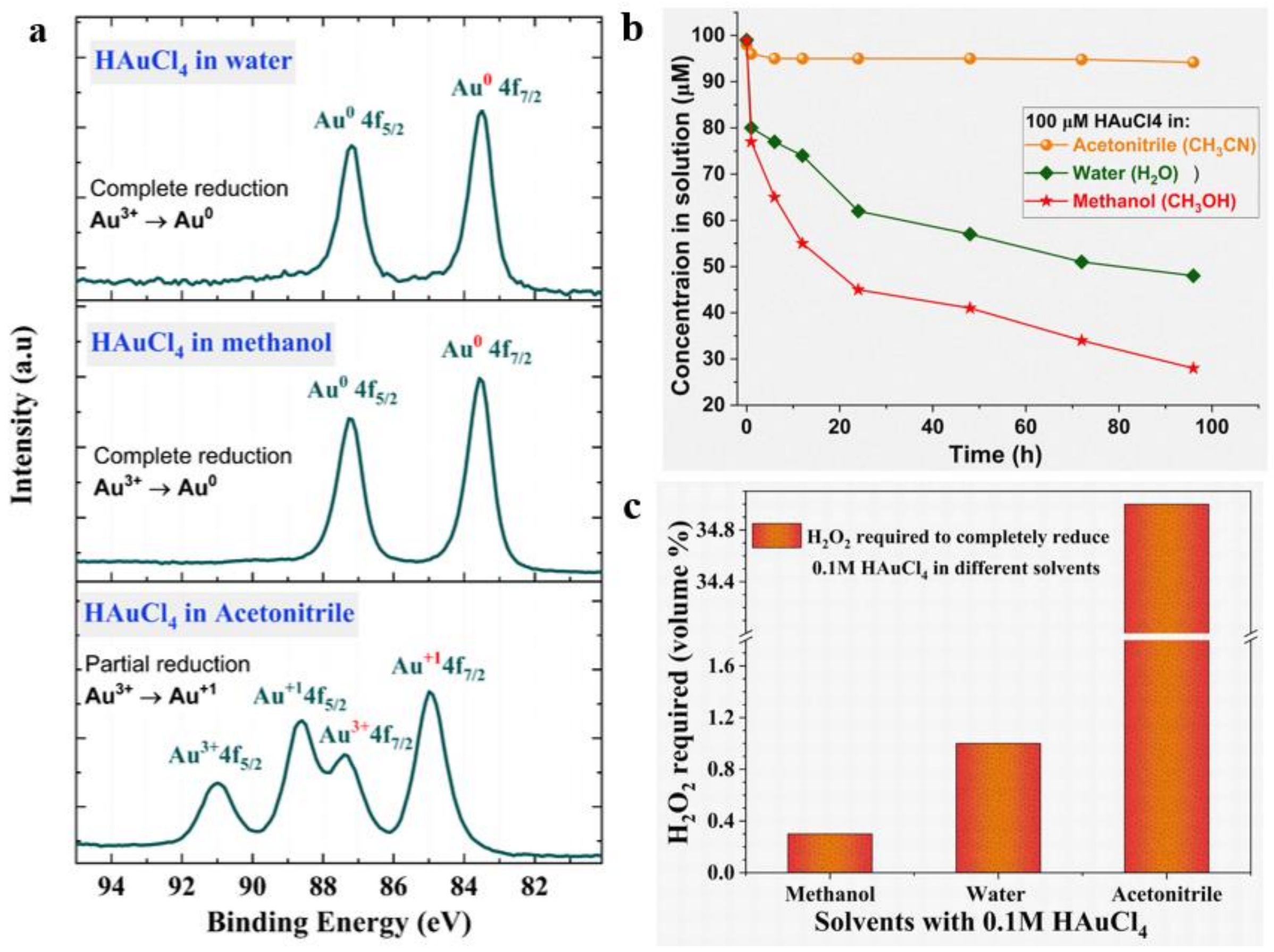

**Fig. 6** Reduction of gold ions in different solvents. **a** Reduction of $Au^{3+}$ in water, methanol, and acetonitrile. **b** Reduction rates of $Au^{3+}$ in water, methanol, and ethanol. **c** $H_2O_2$ required for the complete reduction of 0.1-M $HAuCl_4$ in water, methanol, and acetonitrile. These results suggest that gold ions can undergo spontaneous complete reduction in water and methanol but not in acetonitrile.

Like gold cations, we probed the fate of $Pt^{4+}$ in an aqueous $PtCl_4$ solution and $Fe^{3+}$ in a $FeCl_3$(aq) solution. We found that $Pt^{4+}$(aq) was reduced to Pt(s) NPs and $Pt^{2+}$, whereas $Fe^{3+}$(aq) was reduced only to $Fe^{2+}$(aq) ions, and $Fe^0$(s) was not observed (Figs. 5 and S7). This was because the reduction potential of $Pt^{4+}$ to Pt(s) and $Pt^{2+}$ and that of $Fe^{3+}$ ions to $Fe^{2+}$ were nobler to water oxidation reactions (equations (3) and (4)) (Supplementary Table S2). As the reduction potentials of $Fe^{3+}$ and $Fe^{2+}$ to Fe(s) are −0.037 and −0.44 V, lower than those listed in equations (3) and (4), it is not feasible for $Fe^{3+}$ or $Fe^{2+}$ to reduce to Fe(s) by water. The reduction of $Al^{3+}$, $Cu^{2+}$, $Ni^{2+}$, $Zn^{2+}$, and $Mg^{2+}$ in water also proceeds similarly (Supplementary Table S2).

## Conclusions

In summary, our report refutes the claims that water microdroplets and the air–water interface has unique properties that enable the spontaneous reduction of gold ions. We demonstrated that this chemical transformation was not limited to (i) gold ions, (ii) water as a solvent, and (iii) microdroplets or the air–water interface. In other words, we confirmed that it was a bulk phenomenon. These findings are in alignment with our resolution of the latest controversy on the spontaneous formation of 0.3–1.5 μM of $H_2O_2$(aq) in water microdroplets, where we demonstrated that (i) microdroplets or the air–water interface has no bearing on the phenomenon (detection limit ≥50 nM) and (ii) the $H_2O_2$ formation was due to the reduction of dissolved $O_2$(aq) at the water–solid interface[64, 73]. Essentially, the formation of AuNPs and $H_2O_2$(aq) in water can be explained simply by using the standard electrochemical series. The cations with very high reduction potential, such as $Au^{3+}$ and $Pt^{4+}$, are so unstable that they can oxidize the solvent (water, methanol, and ethanol) to form NPs and $H_2O_2$(aq) (and other byproducts in the case of alcohols). In contrast, when metals with a high propensity for oxidation/corrosion, such as Mg and Al, contact water, they react with the dissolved $O_2$(aq) in water and form $H_2O_2$(aq). In the case of NP formation, the $H_2O_2$(aq) concentration is proportional to that of the metal salt. In contrast, in the case of surface oxidation of metals, the $H_2O_2$(aq) concentration is proportional to dissolved $O_2$(aq). These findings advance our understanding of aquatic chemistry and warrant caution in investigating microdroplet chemistry and its environmental and practical relevance.

## Methods

### Chemicals.

Gold(III) chloride trihydrate ($HAuCl_4.3H_2O$, 520918), platinum(IV) chloride ($PtCl_4$, 206113), iron(III) chloride hexahydrate ($FeCl_3.6H_2O$, 10025-77-1), iron(II) chloride ($FeCl_2$, 37287-0), aluminum(III) chloride ($AlCl_3$, 06220), zinc chloride ($ZnCl_2$, 211273), magnesium chloride hexahydrate ($MgCl_2.6H_2O$, 13152), sodium chloride (NaCl, 7647145), ammonium chloride ($NH_4Cl$, 213330), and copper chloride ($CuCl_2$, 7447394) were used. The following liquids were also used. Deuterium oxide ($D_2O$, 3000007892), methanol-D4+0.03%TMS, acetonitrile-D3 (CAS 2206-26-0), acetonitrile (HPLC grade purchased from Fisher Scientific, batch 1072451), methanol (HPLC LC-MS grade), standard hydrogen peroxide ($H_2O_2$) 30% (Cat.270733), HPLC grade water (Cat. 2594649), and deionized water (DI) obtained from MilliQ Advantage 10 set up (with resistivity 18.2 MΩ cm).

### Characterization of NPs

SEM and TEM analyses were performed to characterize the NPs in the bulk solutions. For SEM, a small drop of liquid containing NPs was drop-coated on a silicon wafer substrate, and the solution was dried at room temperature. TEM images were acquired using a Titan ST Image Corrected (Thermo Fisher) instrument operated at a 300-kV acceleration voltage. EDS spectra and maps were acquired in the STEM mode using a four-quadrant SuperX EDS detector. For sample preparation, a drop of solution was placed on an ultra-thin carbon-coated holey copper grid, blotted with filter paper, and dried under ambient conditions. The presence and growth of NPs were examined via DLS. The growth of Au NPs was observed in 100-µM $HAuCl_4$ bulk solution at varying time scales.

### XPS measurements

For the XPS studies of gold ion reduction in water, 100-µM $HAuCl_4$ solution was used. The drop coating method was employed to prepare the samples using silicon wafers, and Baer's method was closely followed[74]. Kratos Axis Supra instrument equipped with a monochromatic Al Kα X-ray source (hν = 1486.6 eV) operating at a power of 75 W and under UHV conditions in the range of $\sim 10^{-9}$ mbar was used to obtain the data. All the spectra were recorded in hybrid mode, using magnetic and electrostatic lenses and an aperture slot of 300 µm × 700 µm. The high-resolution spectra were acquired at fixed analyzer pass energies of 20 eV. The samples were mounted in a floating mode to avoid differential charging.

**Quantification of $H_2O_2$**

$H_2O_2$ concentration in all the diluted salt solutions was quantified using the hydrogen peroxide assay kit (HPAK). This is based on the principle that when hydrogen peroxide maintains contact with the AbIR peroxidase indicator, it causes fluorescence. Its maximum emission and excitation wavelengths are 674 and 647 nm, respectively. The samples were analyzed by mixing 50 µL of HPAK reaction mixture with 50 µL of samples in a 96-well black/transparent bottom microtiter plate using a SpectraMax M3 microplate reader. For the fluorescence reading, the SoftMax Pro 7 software was used. $H_2O_2$ in the samples was quantified using the calibration curve obtained from the standard samples on the same day.

*Peroxide test strips for semi-quantitative analysis.* For the qualitative estimation of $H_2O_2$ in the aqueous samples, peroxide test strips (Baker Test Strips purchased from VWR International) with a detection limit of 1 ppm were used. These strips use a colorimetric reagent that changes to blue when in contact with $H_2O_2$.

**NMR spectroscopy for $H_2O_2$(aq) quantification**

NMR measurements were performed on the 700 MHz Bruker Avance Neo NMR spectrometer equipped with a 5-mm Z-axis gradient TXO Cryoprobe at 275 K. For water samples, a 6-ms Gaussian 90-degree pulse was applied to excite the proton of hydrogen peroxide, followed by a 50-ms acquisition and a 1-ms recycle delay. For acetonitrile samples, a W5 binomial pulse sequence with gradients or a routine 1D proton pulse was applied for solvent suppression or nonsuppression measurements, respectively. Peak quantification and peak position confirmation of $H_2O_2$ in salt solutions were performed by comparing the results obtained for standard $H_2O_2$ samples.

**GC-MS experiment**

The methanol sample was extracted by adding 2 mL each of water and hexane to initiate phase separation. Approximately 1 µL of the top hexane layer was analyzed using a single quadrupole GC-MS system (Agilent 7890 GC/5975C MSD) coupled to an EI source with an ionization energy of 70 eV. The ion source and mass analyzer temperatures were kept at 230 °C and 150 °C, respectively, with a solvent delay of 0 min. The mass analyzer was tuned according to the manufacturer’s instructions, and the scan was fixed at 15–200 Da. A DB-5MS fused silica capillary column (30 m × 0.25 mm I.D., 0.25-µm film thickness; Agilent J&W Scientific, Folsom, CA) was used for chromatographic separation; the column contained a 5% phenyl and 95% methylpolysiloxane cross-linked stationary phase. Furthermore, helium was the carrier gas at a constant flow rate of 1.0 mL $min^{-1}$. The sample was injected into a 6890A gas chromatograph (Agilent, USA). The oven program was set to 30 °C and held for 10 min; the temperature was ramped at 20 °C/min to 260 °C with a 0-min hold time. The GC inlet temperature was set at 250 °C, and the transfer line temperature to the MS EI source was kept at 320 °C. The sample was injected using an autosampler equipped with a 10-µL syringe into a split/splitless inlet with a split ratio 10:1.

**Oxygen gas analysis**

Headspace analysis of the aqueous $HAuCl_4$ solution was performed using a gas analyzer (Model 310C, SRI, USA) equipped with a thermal conductivity detector (TCD). About 1 mL of the headspace above the sample was withdrawn using a gas-tight syringe (Hamilton, USA) and injected into a 6-foot molecular sieve 13X packed column. The oven temperature of the column was set to 100°C with argon as the carrier gas.

**Acknowledgments**

The co-authors thank Dr. Adair Gallo, Dr. Peng Zhong, and Ms. Nayara Musskopf for building a robust experimental setup in HM's laboratory; Dr. Ryo Mizuta, Scientific Illustrator at KAUST, for preparing illustrations in Fig. 1; Dr. Valentina-Elena Musteata, scientist from IAC at KAUST for her assistance with TEM; Mohammed Abdulwahab ElShaerfrom from WDRC at Kaust for his assistance with DLS and Usman Sharif from Laboratory Equipment Maintenance (LEM) team at Kaust for his contribution to maintain the XPS instrument at the best operating conditions.

**Author contributions:** ME and HM designed the experiments, which ME performed. NK and ME performed LC-MS, GC, and ICP-OES experiments. XG and ME performed NMR spectroscopy experiments. NW and ME performed XPS studies. HM and ME wrote the manuscript.

**Competing interests:** The authors declare no competing interests.

**Data and materials availability:** All data needed to evaluate the conclusions in the paper are present in the paper or the Supplementary Materials.

**Funding:** HM acknowledges KAUST for funding (Grant No. BAS/1/1070-01-01).

10.1116/1.5141419.

**Supplementary Information**

# Why Some Metal Ions Spontaneously Form Nanoparticles in Water Microdroplets: Disentangling the Contributions of the Air–Water Interface and Bulk Redox Chemistry

Muzzamil Ahmad Eatoo[a,b,c,d], Nimer Wehbe[e], Najeh Kharbatia[b], Xianrong Guo[e], Himanshu Mishra[a,b,c,d*]

[a]Environmental Science and Engineering (EnSE) Program, Biological and Environmental Science and Engineering (BESE) Division, Water Desalination and Reuse Center (WDRC)

[b]Water Desalination and Reuse Center (WDRC),

King Abdullah University of Science and Technology (KAUST), Thuwal, 23955-6900, Kingdom of Saudi Arabia

[c]Center for Desert Agriculture (CDA), King Abdullah University of Science and Technology (KAUST), Thuwal 23955-6900, Saudi Arabia

[d]Interfacial Lab (iLab), King Abdullah University of Science and Technology (KAUST), Thuwal 23955-6900, Saudi Arabia

[e]Core Labs, King Abdullah University of Science and Technology (KAUST), Thuwal 23955-6900, Saudi Arabia

*Correspondence: himanshu.mishra@kaust.edu.sa

## Section S1: Water microdroplet generation via sprays

We adapted the experimental setup built by Gallo et al.[1] to produce water microdroplets. In a coaxial system, water was injected through an inner tube with a 100-µm diameter using a syringe pump (PHD Ultra, Harvard Apparatus). Dry $N_2$(g) was pushed through the outer tube with a 430-µm diameter. Additionally, HPLC-grade water was used to make salt solutions, and a glass cell (equipped with a tiny opening to prevent pressure build-up) was employed to collect microdroplets while minimizing ambient contamination. The water flow rate was 25 µL/min, whereas the gas ($N_2$) pressure was 100 psi.

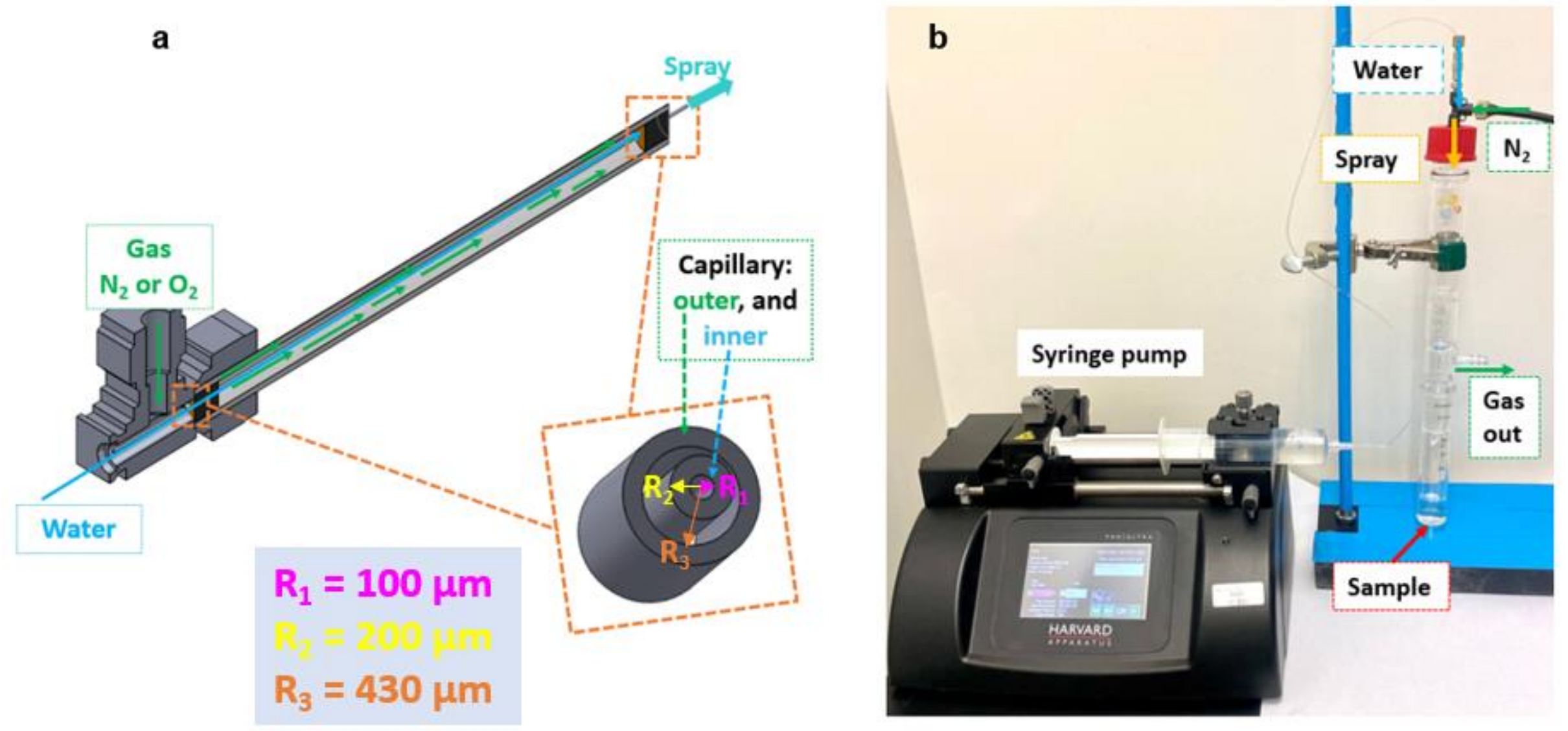

**Fig. S1**: **a)** Schematics of pneumatic spray setup. **b)** Photograph of spray setup connected to a removable glass flask for sample collection, comprising two coaxial capillary tubes. Liquid flows through the inner tube, whereas nitrogen gas flows through the outer annulus. Metal salt solution samples were injected via a syringe pump, and a high-pressure cylinder supplied $N_2$ gas. Sprayed microdroplets were collected in a clean glass flask until a sufficient analyte volume was collected.[2]

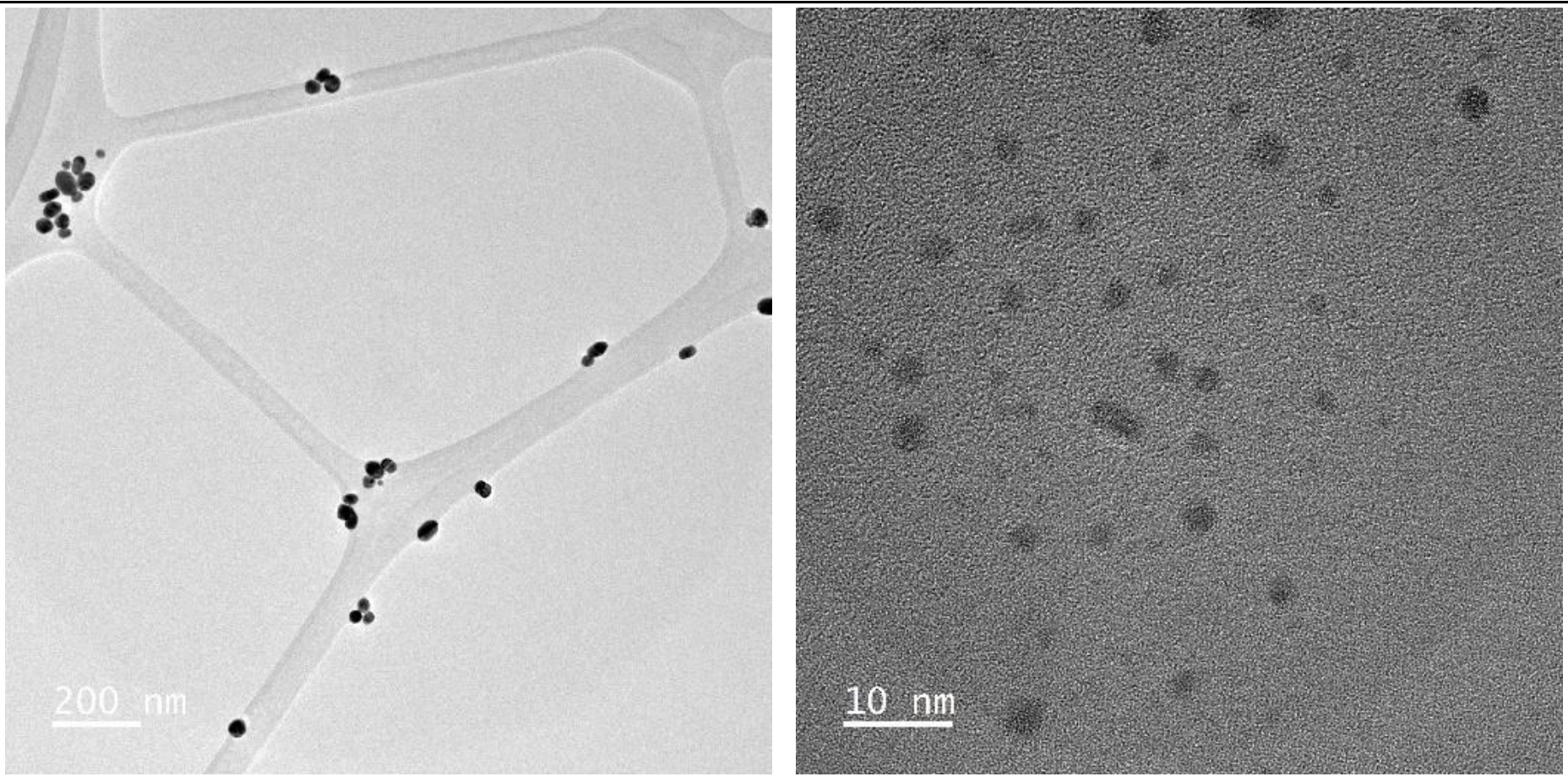

**Fig. S2** TEM imaging of microdroplets samples formed by spraying 100 µM $HAuCl_4$ solution, collected and drop cast, shows the presence of AuNPs.

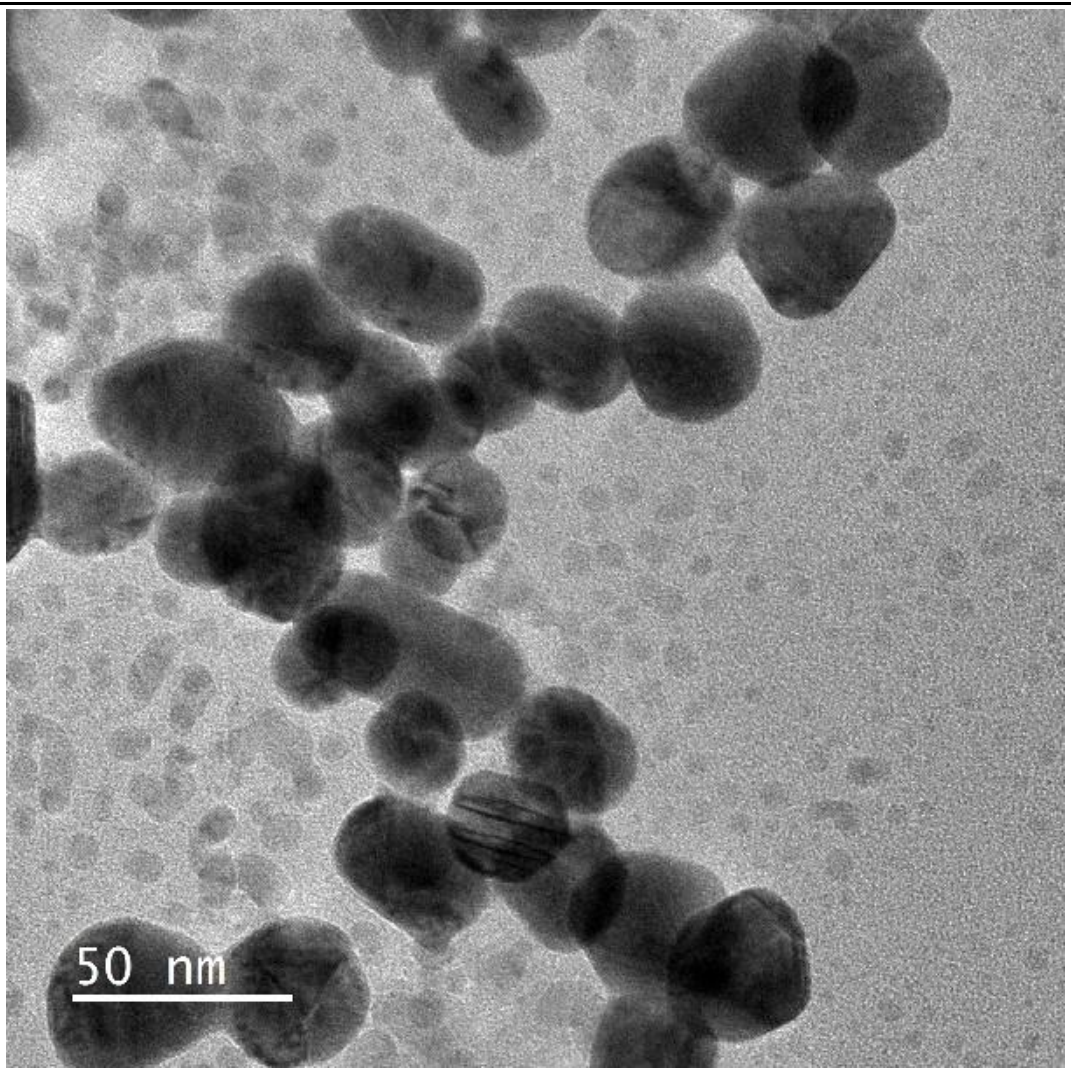

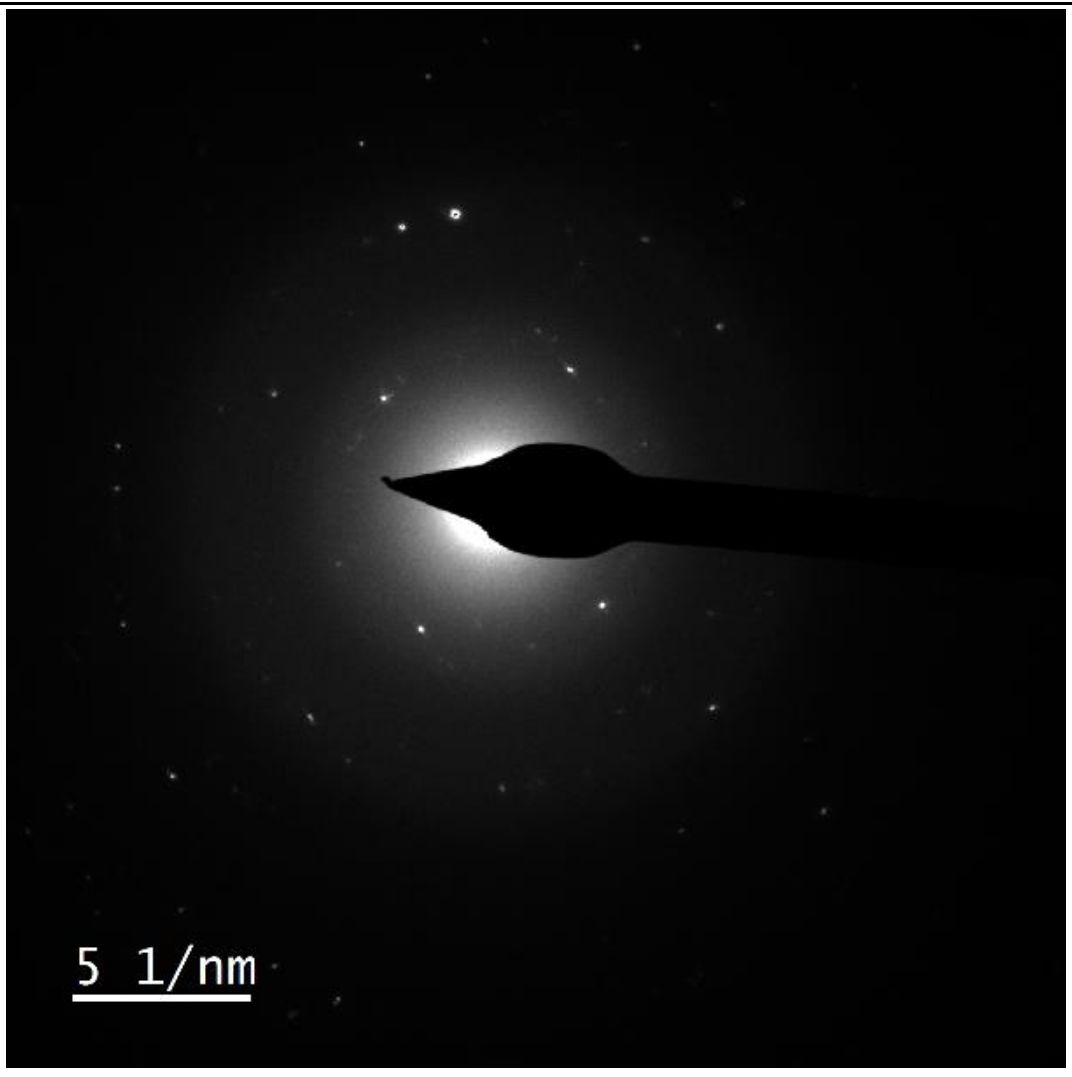

**Fig. S3** TEM imaging of bulk aqueous 100 µM $HAuCl_4$ solution: Micrograph showing the presence of nanoparticles of different sizes and diffraction pattern showing the formation of Au(0).

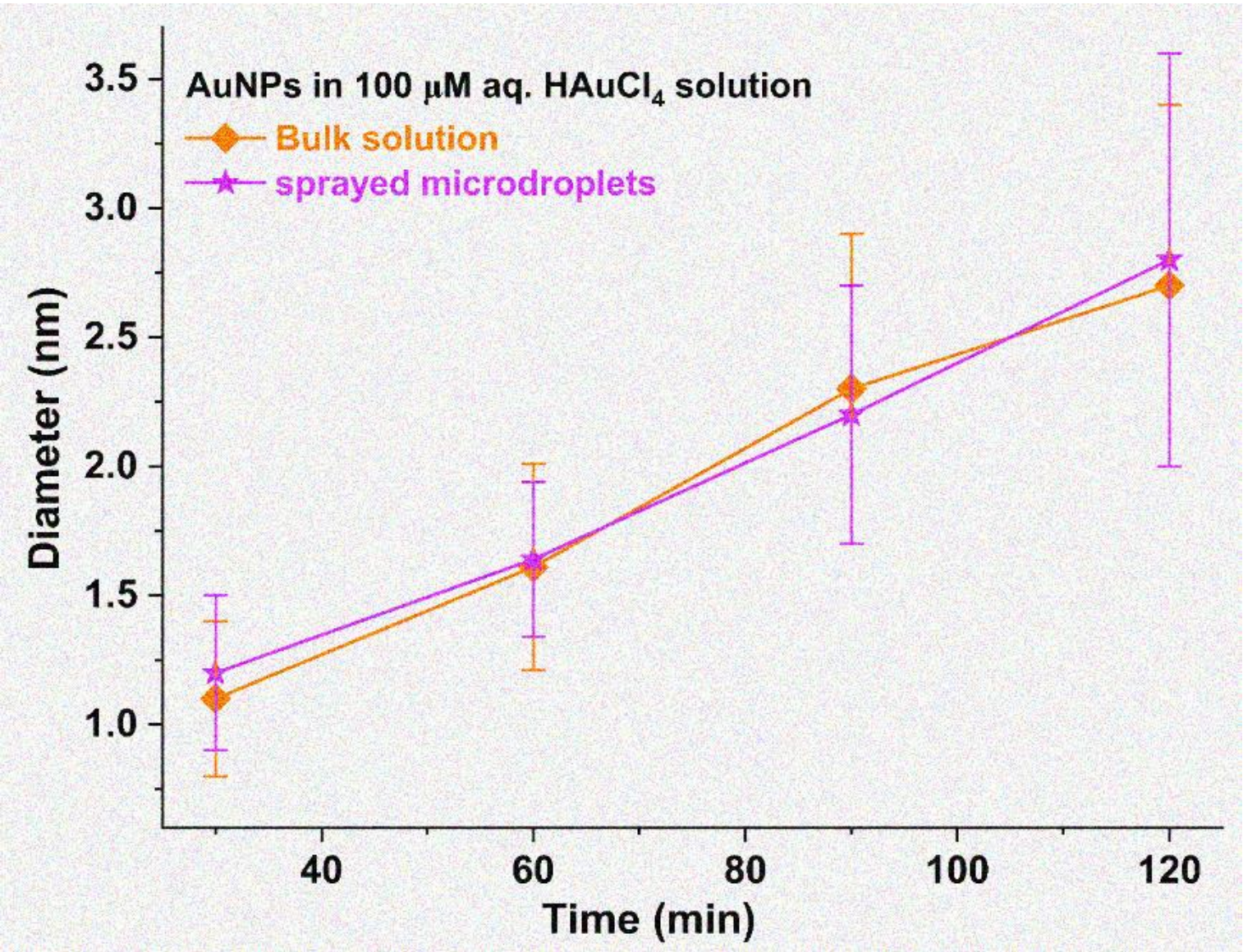

**Fig. S4** shows DLS measurements for the growth kinetics of nanoparticles in microdroplets vs. bulk solution.

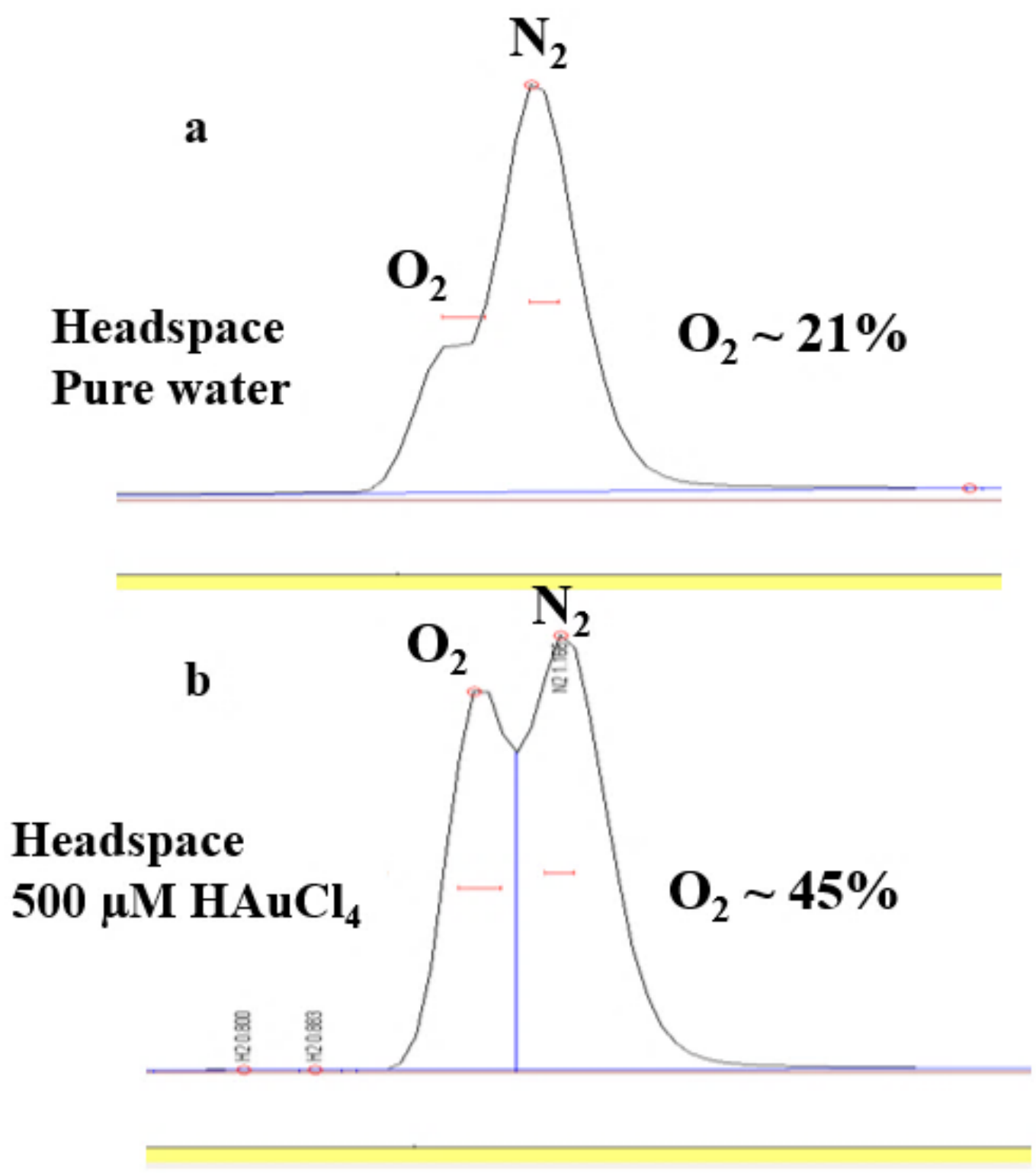

**Fig. S5**: Gas chromatography results show oxygen gas evolution from the $HAuCl_4$ aqueous solution. **a** showing the presence of around 21% oxygen in the headspace of pure water (DI used for making solutions). **b** shows the presence of around 45% oxygen in the headspace of 500 µM $HAuCl_4$ solution after the age of around 24 h. This reveals that the $HAuCl_4$ solution evolves oxygen gas.

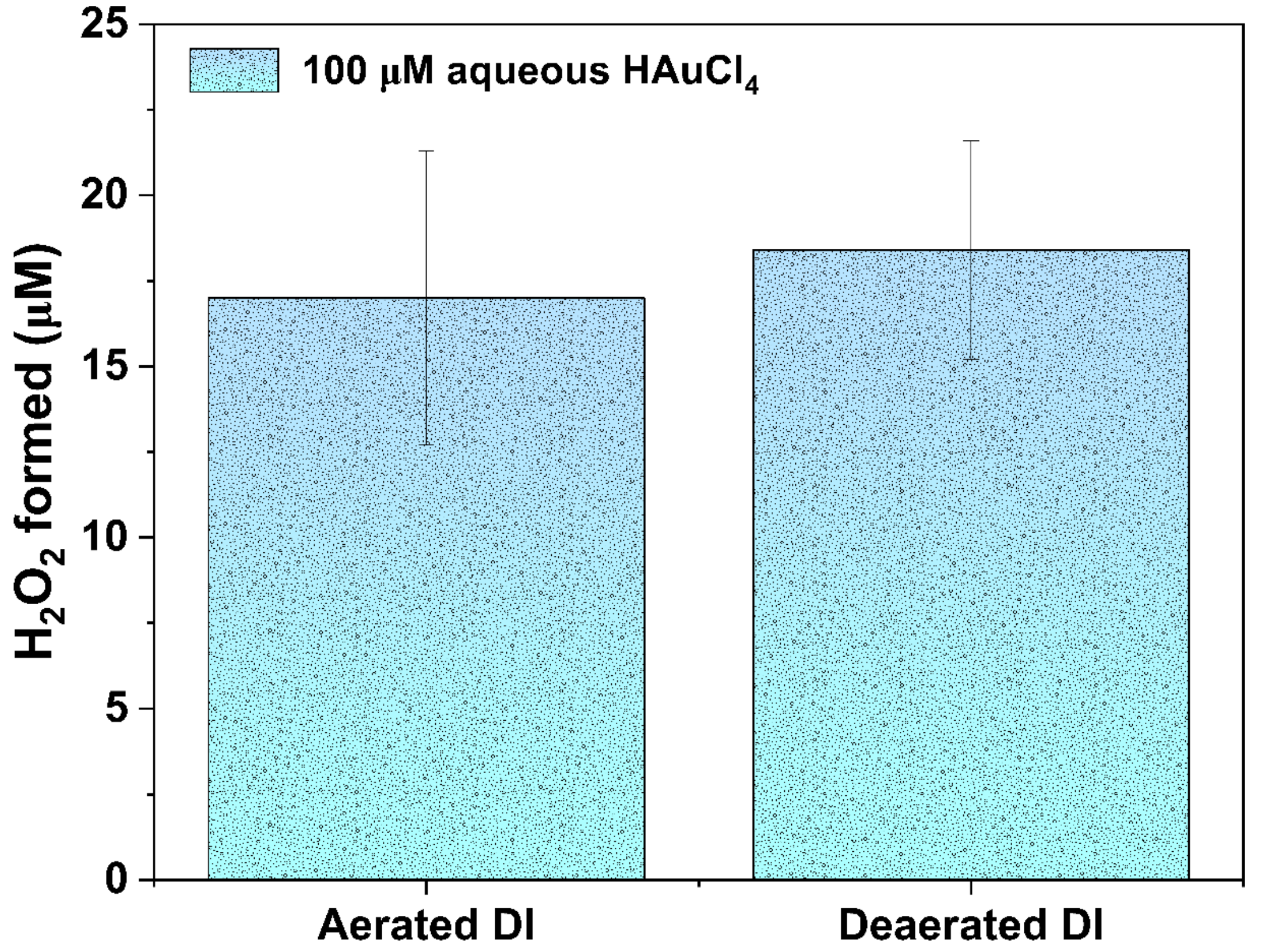

**Fig. S6** Effect of water deaeration on $H_2O_2$ formation by $Au^{3+}$ cations in water. These results reveal that deaeration has no significant effect on the $H_2O_2$ formation.

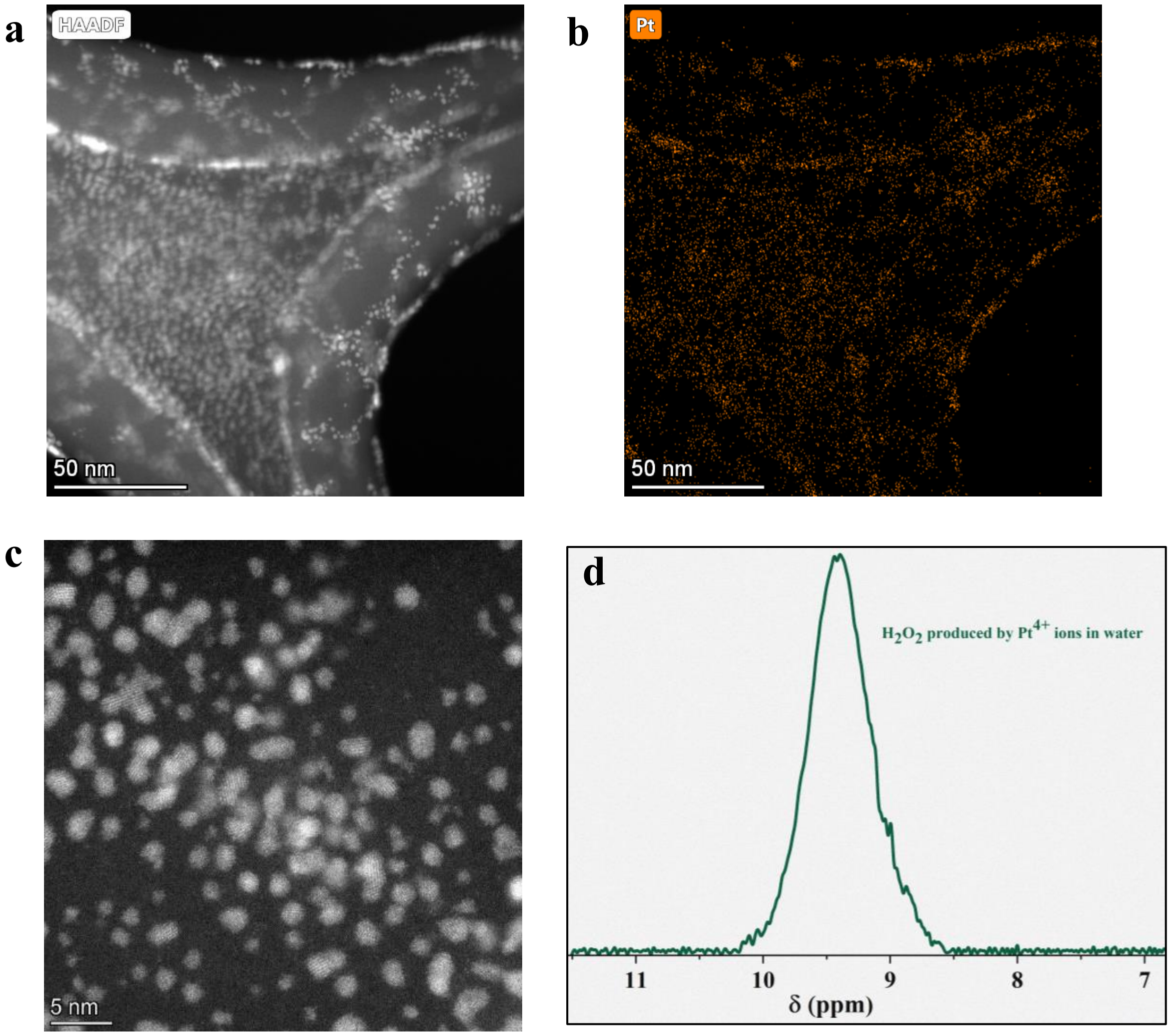

**Fig. S7** Characterization of bulk aqueous PtCl4 solution. **a-c** TEM imaging of drop casted bulk $PtCl_4$ solution shows the presence of Pt nanoparticles. **d** NMR spectroscopy results revealing $H_2O_2$ formation by $Pt^{4+}$ ions in water.

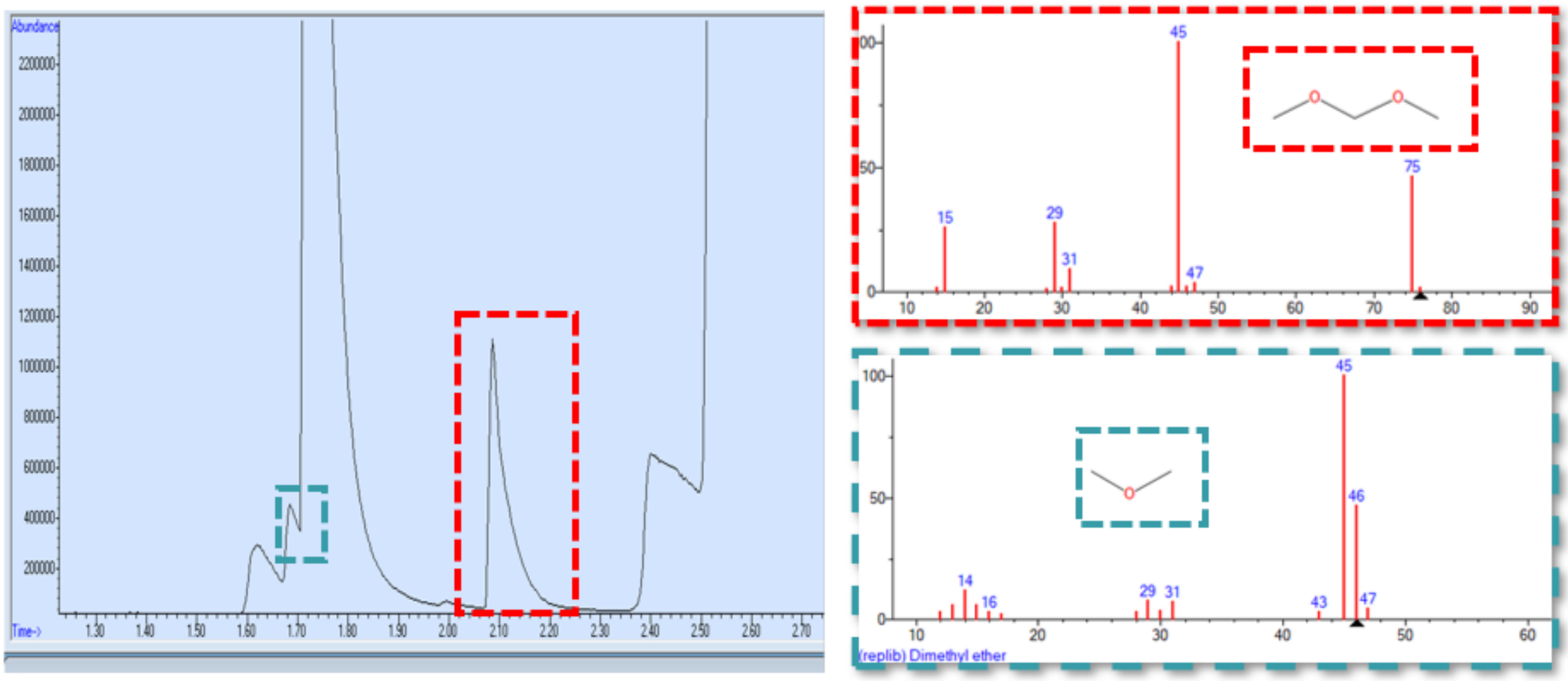

**Fig. S8** GC-MS for $HAuCl_4$ salt in methanol. The results reveal the presence of Methylal ($CH_3OCH_2OCH_3$) and Dimethyl ether ($CH_3OCH_3$)

**Table S1** Summary of some key observations in this study

| Solvent | Environment (bulk or microdroplet) | Formation of Nanoparticles $M^{n+} + ne^- = M(s)$ | Metal ions which form $H_2O_2$ | Kinetics of Formation of Nanoparticles $M^{n+} + ne^- = M(s)$ |
|---|---|---|---|---|
| Water | Bulk | Au and Pt nanoparticles | $Au^{3+}$, $Pt^{4+}$, $Fe^{3+}$ | Fast |
| | Microdroplet | | | |
| Methanol | Bulk | Au and Pt nanoparticles | $Au^{3+}$, $Pt^{4+}$, $Fe^{3+}$ | Fast |
| Ethanol | Bulk | Au and Pt nanoparticles | $Au^{3+}$, $Pt^{4+}$, $Fe^{3+}$ | Fast |
| Acetonitrile | Bulk | Negligible NPs formation | No $H_2O_2$ | Sluggish |

**Table S2**:

Standard electrochemical series with the standard reduction potentials:

| Standard Reduction Half-Reaction | Standard Reduction Potential E° (volts) |
|---|---|
| | |
| $Au^{+} + e^{-} \rightleftharpoons Au(s)$ | 1.83 |
| $Au^{3+} + 3e^{-} \rightleftharpoons Au(s)$ | 1.52 |
| $Au^{3+} + 2e^{-} \rightleftharpoons Au^{+}$ | 1.36 |
| $AuCl_4^{-} + 3e^{-} \rightleftharpoons Au(s) + 4Cl^{-}$ | 1.002 |
| $AuCl_4^{-} + 2e^{-} \rightleftharpoons AuCl_2^{-} + 2Cl^{-}$ | 0.926 |
| | |
| $Pt^{2+} + 2e^{-} \rightleftharpoons Pt(s)$ | 1.2 |
| $PtCl_2 + 2e^{-} \rightleftharpoons Pt(s) + 4Cl^{-}$ | 0.78 |
| $PtCl_4 + 2e^{-} \rightleftharpoons PtCl_2 + 2Cl^{-}$ | 0.75 |
| $Fe^{3+} + e^{-} \rightleftharpoons Fe^{2+}$ | 0.771 |
| **$H_2O_2 + H^{+} + e^{-} \rightleftharpoons HO^{*} + H_2O$** | **0.710** |
| $O_2(g) + 2H^{+} + 2e^{-} \rightleftharpoons H_2O_2$ | 0.695 |
| **$O_2(g) + 2H_2O(l) + 4e^{-} \rightleftharpoons 4OH^{-}$** | **0.401** |
| $Cu^{2+} + 2e^{-} \rightleftharpoons Cu$ | 0.34 |
| $Fe^{3+} + 3e^{-} \rightleftharpoons Fe(s)$ | -0.037 |
| $Ti^{2+}+2e^{-}\rightleftharpoons Ti(s)$ | -0.163 |
| $Ti^{3+} + e^{-} \rightleftharpoons Ti^{2+}$ | -0.37 |
| $Fe^{2+} + 2e^{-} \rightleftharpoons Fe(s)$ | -0.44 |
| $Zn^{2+}(aq) + 2e^{-} \rightleftharpoons Zn(s)$ | -0.76 |
| $Al^{3+} + 3e^{-} \rightleftharpoons Al(s)$ | -1.706 |
| $Mg^{2+} + 2e^{-} \rightleftharpoons Mg(s)$ | -2.356 |

**Section S2:** Confirmation of $H_2O_2$ peak position

To confirm that the obtained peak positioned at a chemical shift around 9.4 ppm in gold ion solution (0.3M $HAuCl_4$ in water) corresponds to the $H_2O_2$ only (Fig. S9), the peak was compared to one obtained for the same gold ion concentration (0.3M $HAuCl_4$) with the addition of 200µM $H_2O_2$ (i.e., 0.3M $HAuCl_4$ + 200 µM $H_2O_2$) and the new peak was found at the same position/chemical shift with relatively more intensity and no other additional peak/s was found in the spectra. This confirms that the peak at around 9.4 ppm corresponds to $H_2O_2$ only.

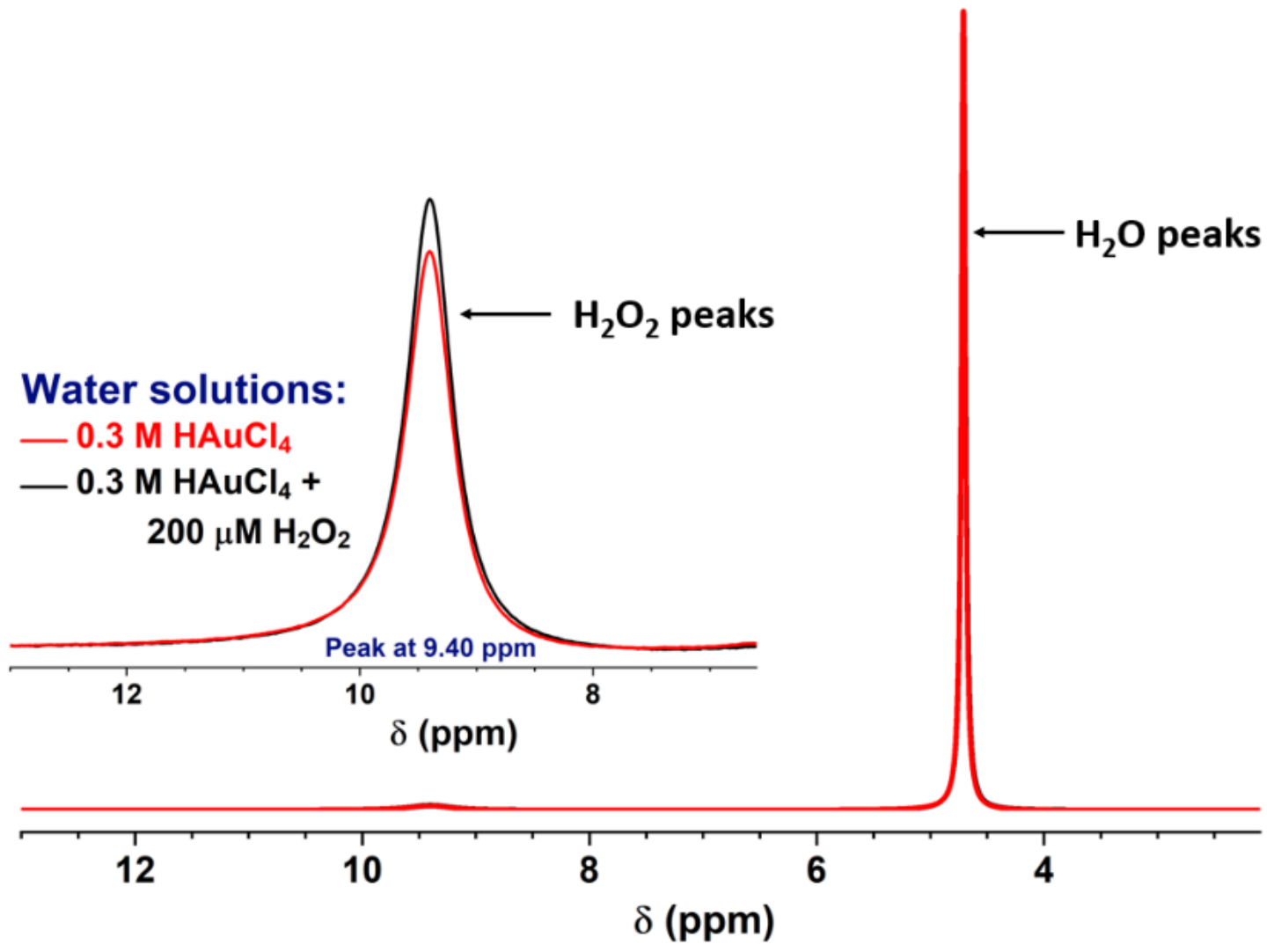

**Fig. S9** NMR Spectra confirms the peak position of $H_2O_2$ in water produced by $Au^{3+}$ions.

**Fig. S10** shows the data obtained for the same concentration of gold ion ($Au^{3+}$) in water and acetonitrile, which confirmed the formation of $H_2O_2$ in water and not in acetonitrile.

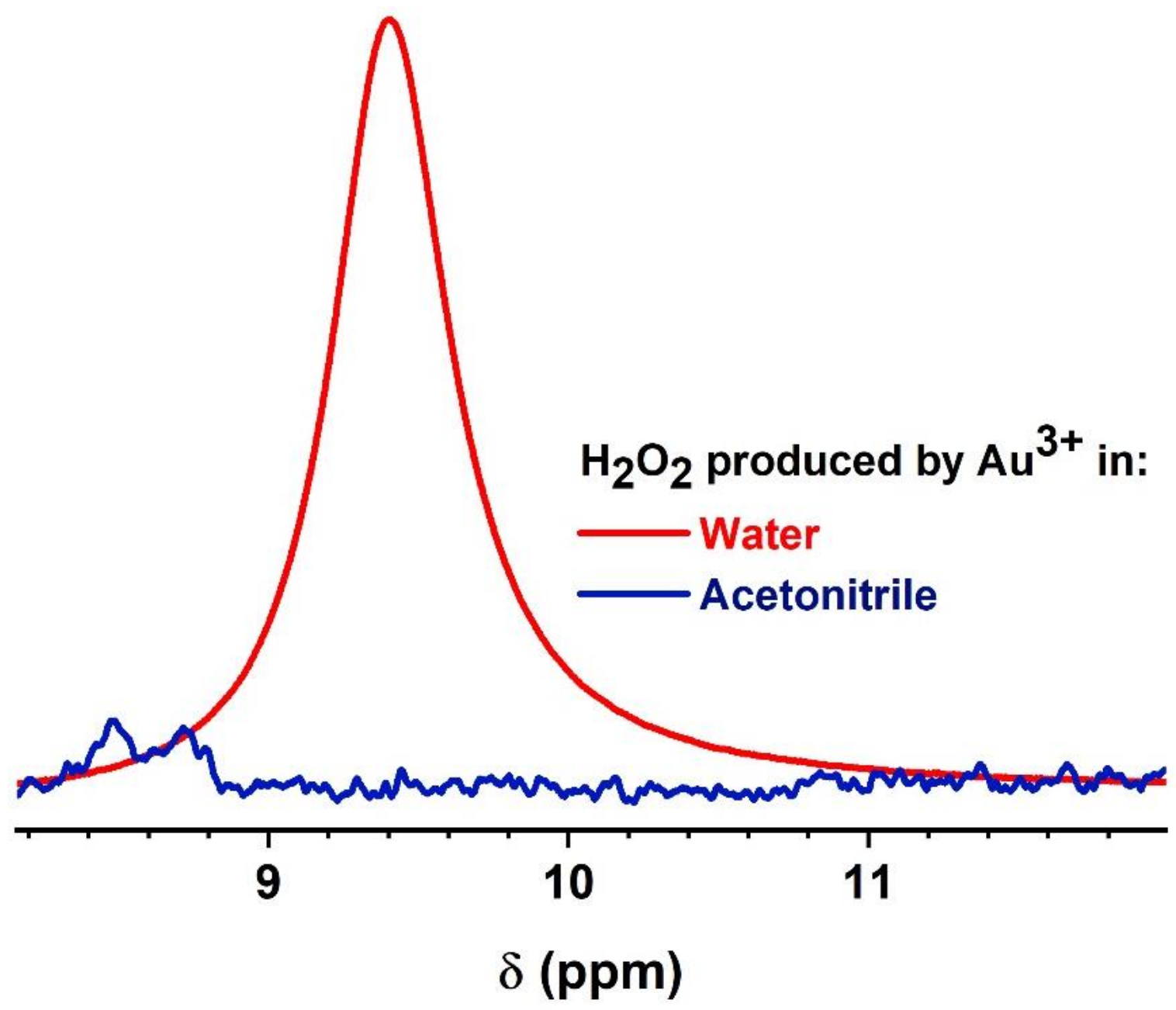

**Fig. S10** NMR Spectra showing that $Au^{3+}$ produces $H_2O_2$ in water and not in acetonitrile.